\documentclass[english,conference, two column]{IEEEtran}
\usepackage[T1]{fontenc}
\usepackage[latin9]{inputenc}
\usepackage{verbatim}
\usepackage{textcomp}
\usepackage{amsthm}
\usepackage{amsmath}
\usepackage{amssymb}
\usepackage{graphicx}

\makeatletter

\newcommand{\lyxmathsym}[1]{\ifmmode\begingroup\def\b@ld{bold}
  \text{\ifx\math@version\b@ld\bfseries\fi#1}\endgroup\else#1\fi}

\theoremstyle{definition}
\newtheorem{property}{Property}
  \theoremstyle{plain}
  \newtheorem{thm}{\protect\theoremname}
  \theoremstyle{remark}
  \newtheorem{rem}{\protect\remarkname}
  \theoremstyle{plain}
  \newtheorem{prop}{\protect\propositionname}
\theoremstyle{definition}
\newtheorem{assumption}{Assumption}

\IEEEoverridecommandlockouts
\allowdisplaybreaks
\usepackage{cite}

\makeatother

\usepackage{babel}
\providecommand{\propositionname}{Proposition}
\providecommand{\remarkname}{Remark}
\providecommand{\theoremname}{Theorem}

\begin{document}

\title{Stationary Cycling Induced by Switched Functional Electrical Stimulation
Control%
\thanks{1. Department of Mechanical and Aerospace Engineering, University
of Florida, Gainesville FL 32611-6250, USA Email: \{mattjo, tenghu,
ryan2318, wdixon\}@ufl.edu%
}%
\thanks{Research for this paper was conducted with Government support under
FA9550-11-C-0028 and awarded by the Department of Defense, Air Force
Office of Scientific Research, National Defense Science and Engineering
Graduate (NDSEG) Fellowship, 32 CFR 168a.%
}%
\thanks{This research is supported in part by NSF award number 1161260. Any
opinions, findings and conclusions or recommendations expressed in
this material are those of the authors and do not necessarily reflect
the views of the sponsoring agency.%
}}

\author{M. J. Bellman$^{1}$, T. -H. Cheng$^{1}$, R. J. Downey$^{1}$, and
W. E. Dixon$^{1}$}
\maketitle
\begin{abstract}
Functional electrical stimulation (FES) is used to activate the dysfunctional
lower limb muscles of individuals with neuromuscular disorders to
produce cycling as a means of exercise and rehabilitation. %
In this paper, a stimulation pattern for quadriceps femoris-only FES-cycling
is derived based on the effectiveness of knee joint torque in producing
forward pedaling. In addition, a switched sliding-mode controller
is designed for the uncertain, nonlinear cycle-rider system with autonomous
state-dependent switching. The switched controller yields ultimately
bounded tracking of a desired trajectory in the presence of an unknown,
time-varying, bounded disturbance, provided a reverse dwell-time condition
is satisfied by appropriate choice of the control gains and a sufficient
desired cadence. Stability is derived through Lyapunov methods for
switched systems, and experimental results demonstrate the performance
of the switched control system under typical cycling conditions.
\end{abstract}

\section{Introduction}

Since the 1980s, cycling induced by functional electrical stimulation
(FES) has been investigated as a safe means of exercise and rehabilitation
for people with lower-limb paresis or paralysis \cite{Phillips.Petrofsky.ea1984},
and numerous physiological and psychological benefits have since been
reported \cite{Peng.Chen.ea2011}. Despite these benefits, FES-cycling
is still metabolically inefficient and results in low power output
compared to volitional cycling by able-bodied individuals \cite{Hunt.Fang.ea2012},
\cite{Hunt.Hosmann.ea2012}. Many approaches have been taken in attempt
to improve the efficiency and power output of FES-cycling, including
optimization of the cycling mechanism \cite{Szecsi.Krause.ea2007},
\cite{Ibrahim.Gharooni.ea2008}, alteration of the stimulation pattern
\cite{Gfohler.Lugner2000,Idso.Johansen.ea2004,Trumbower.Faghri2004,Hunt.Ferrario.ea2006,Hakansson.Hull2009},
and variation of the stimulation strategy \cite{Perkins.Donaldson.ea2002,Eser.Donaldson.ea2003,Decker.Griffin.ea2010,Hakansson.Hull2012}.
Studies have explored the use of feedback control methods to automatically
determine the appropriate stimulation parameters, but most use either
linear approximations of the nonlinear cycle-rider system \cite{Hunt.Stone.ea2004}
or nonlinear methods lacking detailed stability analyses \cite{Chen.Yu.ea1997,Kim.Eom.ea2008,Farhoud.Erfanian2014}.%
{} Previous approaches (cf. \cite{Pons.Vaughan.ea1989}, \cite{Petrofsky2003})
often either determine the stimulation pattern manually by stimulating
a particular muscle group and observing the resulting crank motion
or base the stimulation pattern on electromyography recordings of
able-bodied cyclists. Stimulation frequency is then fixed and stimulation
intensity is typically controlled by varying the magnitude of a predefined
trapezoidal stimulation signal proportional to the cycling cadence
error. All of the aforementioned studies used stimulation patterns
which switch between active muscle groups according to the crank position,
resulting in a switched control system with autonomous%
\footnote{Autonomous in the context of switched systems means the switching
happens automatically and is not manually controlled by an actor.%
} state-dependent switching \cite{Liberzon2003}. However, no study
has yet investigated FES-cycling control in the context of switched
systems theory. Switched systems may be unstable even if each subsystem
is asymptotically stable \cite{Liberzon2003}, so the switching logic
must be taken into account in the control design to guarantee stability
of the system. In general, switched control of FES-cycling involves
switching between stabilizable and unstable, uncertain, nonlinear
dynamics, yet no previous studies have addressed this aspect of FES-cycling.
Therefore, studying FES-cycling in this light could potentially reveal
new control methods. 

In this paper, a nonlinear model of the cycle-rider system is considered
with parametric uncertainty and an unknown, time-varying, bounded
disturbance, and the stability of the closed-loop switched control
system is analyzed via Lyapunov methods. A stimulation pattern for
the quadriceps femoris muscle groups of the rider is derived based
on the kinematic effectiveness of knee joint torque at producing forward
pedaling throughout the crank cycle. The stimulation pattern is designed
so that stimulation is not applied in regions near the dead points
of the cycle to bound the stimulation voltage, while everywhere else
in the crank cycle either the right or left quadriceps are stimulated
to produce forward pedaling. Since the quadriceps are stimulated without
overlap, the crank cycle is composed of controlled and uncontrolled
regions. Stimulation of additional muscle groups, e.g., the gluteal
muscles, could eliminate the uncontrolled regions but it could also
result in overlapping controlled regions and over-actuation; thus,
only stimulation of the quadriceps is considered in this paper. In
the controlled regions, a switched sliding-mode controller for the
stimulation intensity is designed which guarantees exponentially stable
tracking of the desired crank trajectory provided sufficient gain
conditions are satisfied. In the uncontrolled regions, the tracking
error is proven to be bounded provided a reverse dwell-time condition
is satisfied, which requires that the system must not dwell in the
uncontrolled region for an overly long time interval \cite{Hespanha.Liberzon.ea2008}.
The reverse dwell-time condition is shown to be satisfied provided
sufficient desired cadence conditions are satisfied. The tracking
error is then proven to converge to an ultimate bound as the number
of crank cycles approaches infinity. Experimental results are provided
which demonstrate the performance of the switched controller for typical
cycling conditions.

\section{Model}

\subsection{Stationary Cycle and Rider Dynamic Model}

The subsequent development is based on a stationary cycle and a two-legged
rider that are modeled as a single degree-of-freedom system \cite{Idso2002},
which can be expressed as
\begin{equation}
M\ddot{q}+V\dot{q}+G+\tau_{d}-\tau_{b}-P=\sum_{s\in\mathcal{S}}B_{k}^{s}\Omega^{s}u^{s},\label{eq:Final General Dynamics Expression}
\end{equation}
where $q\in\mathcal{Q}\subseteq\mathbb{R}$ denotes the crank angle,
defined as the clockwise angle between the ground and the right crank
arm, $M\in\mathbb{R}$ denotes inertial effects, $V\in\mathbb{R}$
represents centripetal and Coriolis effects, $G\in\mathbb{R}$ represents
gravitational effects, $\tau_{d}\in\mathbb{R}$ represents an unknown,
time-varying, bounded disturbance (e.g., changes in load), $\tau_{b}\triangleq-c\dot{q}$
represents viscous damping in the crank joint bearings where $c\in\mathbb{R}_{>0}$
is the unknown constant damping coefficient, $P\in\mathbb{R}$ captures
the passive viscoelastic effects of the rider's joints on the crank's
motion, $B_{k}\in\mathbb{R}$ is the Jacobian element relating torque
about the knee to torque about the crank, and $\Omega\in\mathbb{R}$
is an uncertain nonlinear function relating the quadriceps stimulation
voltage $u\in\mathbb{R}$ to the active torque at the knee. The superscript
$s\in\mathcal{S}\triangleq\left\{ R,\: L\right\} $ denotes right
($R$) and left ($L$) sides of the model (i.e., right and left legs
and crank arms) and is omitted unless it adds clarity. The rider's
legs are modeled as planar rigid-body segments with revolute hip and
knee joints (more complex models of the knee joint have a negligible
effect on the linkage kinematics \cite{Schutte.Rodgers.ea1993}).
The ankle joint is assumed to be fixed in accordance with common clinical
cycling practices for safety and stability \cite{Gfohler.Lugner2000}.
When the rider's feet are fixed to the pedals, the resulting system's
position and orientation can be completely described by the crank
angle (or any other single joint angle measured with respect to ground),
the kinematic parameters of the limb segments, and the horizontal
and vertical components of the distance between the axes of rotation
of the crank and hip joints, $l_{x},\: l_{y}\in\mathbb{R}_{\geq0},$
respectively. The model in (\ref{eq:Final General Dynamics Expression})
has the following properties\cite{Sharma2009,Ferrarin2000,Schauer2005}.%

\begin{property}
$c_{m}\leq M\leq c_{M},$ where $c_{m},\: c_{M}\in\mathbb{R}_{>0}$
are known constants. \textbf{Property 2.} $\left|V\right|\leq c_{V}\left|\dot{q}\right|$,
where $c_{V}\in\mathbb{R}_{>0}$ is a known constant. \textbf{Property
3.} $\left|G\right|\leq c_{G},$ where $c_{G}\in\mathbb{R}_{>0}$
is a known constant. \textbf{Property 4. }$\left|\tau_{d}\right|\leq c_{d},$
where $c_{d}\in\mathbb{R}_{>0}$ is a known constant. \textbf{Property
5. }$\left|B_{k}^{s}\right|\leq c_{B}$ $\forall s\in\mathcal{S},$
where $c_{B}\in\mathbb{R}_{>0}$ is a known constant. \textbf{Property
6. }$\left|P\right|\leq c_{P1}+c_{P2}\left|\dot{q}\right|$, where
$c_{P1},\: c_{P2}\in\mathbb{R}_{>0}$ are known constants. \textbf{Property
7. }$c_{\Omega1}\leq\Omega^{s}\leq c_{\Omega2}\:\forall s\in\mathcal{S},$
where $c_{\Omega1},\: c_{\Omega2}\in\mathbb{R}_{>0}$ are known constants.
\textbf{Property 8. }$\frac{1}{2}\dot{M}-V=0.$
\end{property}

\subsection{Switched System Model\label{sec:Switched-System-Model} }

The torque transfer ratios $B_{k}^{s}$ give insight into how the
quadriceps muscles of each leg should be activated during the crank
cycle. By convention, the quadriceps can only produce a counter-clockwise
torque about the knee joint that acts to extend the knee. Multiplication
of the active knee torque by $B_{k}$ transforms the counter-clockwise
torque produced by the quadriceps to a resultant torque about the
crank. Therefore, since forward pedaling requires a clockwise torque
about the crank, the quadriceps muscles should only be activated when
they produce a clockwise torque about the crank%
, i.e., when $B_{k}$ is negative. The torque transfer ratio $B_{k}$
is negative definite for half of the crank cycle and the sign of the
torque transfer ratio of one leg is always opposite that of the other
leg (i.e., $B_{k}^{R}B_{k}^{L}\leq0\:\forall q$), provided the crank
arms are offset by $\pi$ radians. To induce pedaling using only FES
of the quadriceps muscles, a controller must stimulate muscles on
each leg in an alternating pattern, using the right quadriceps when
$B_{k}^{R}<0$, the left quadriceps when $B_{k}^{L}<0$, and switching
between muscle groups when $B_{k}=0.$ The torque transfer ratios
$B_{k}^{s}$ are zero only at the so-called dead points $q^{*}\in\mathcal{Q}^{*}\subset\mathcal{Q}$,
where $\mathcal{Q}^{*}\triangleq\left\{ q\in\mathcal{Q}\:|\: q=\mbox{arctan}\left(l_{y}/l_{x}\right)+i\pi\right\} ,\: i\in\mathbb{Z}.$

Since the torque transfer ratios are minimal near the dead points,
stimulation applied close to the dead points is inefficient in the
sense that large knee torques yield small crank torques. Therefore,
in the subsequent development, the uncontrolled region in which no
stimulation is applied is defined as $\mathcal{Q}_{u}\triangleq\left\{ q\in\mathcal{Q}\:|\:-B_{k}\left(q\right)\leq\varepsilon\right\} \subset\mathcal{Q},$
where $\varepsilon\in\mathbb{R}_{>0}$ is a scalable constant that
can be increased to reduce the portion of the cycle trajectory where
stimulation is applied. Specifically, $\varepsilon<\mbox{max}\left(-B_{k}\right).$
Also let the sets $\mathcal{Q}^{s}\subset\mathcal{Q}$ be defined
as the regions where the right and left quadriceps muscle groups are
stimulated, as
\begin{equation}
\mathcal{Q}^{s}\triangleq\left\{ q\in\mathcal{Q}\:|\:-B_{k}^{s}\left(q\right)>\varepsilon\right\} ,\label{eq: controlled regions}
\end{equation}
and denote the set $\mathcal{Q}_{c}\triangleq\cup_{s\in\mathcal{S}}\mathcal{Q}^{s}$
as the controlled region, where $\mathcal{Q}_{c}\cup\mathcal{Q}_{u}=\mathcal{Q}$
and $\mathcal{Q}^{R}\cap\mathcal{Q}^{L}=\textrm{Ø},$ i.e., stimulation
voltage is never applied to both legs at the same time. The stimulation
pattern is then completely defined by the cycle-rider kinematics (i.e.,
$B_{k}$) and selection of $\varepsilon.$ Smaller values of $\varepsilon$
yield larger stimulation regions and vice versa. Some evidence in
the FES-cycling literature (e.g., \cite{Idso.Johansen.ea2004}) suggests
that the stimulation region should be made as small as possible while
maximizing stimulation intensity to optimize metabolic efficiency,
motivating the selection of large values of $\varepsilon.$

\begin{figure}
\centering{}\includegraphics[width=0.8\columnwidth]{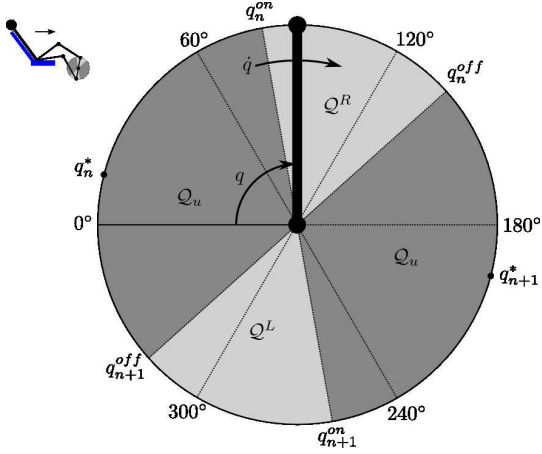}\caption{Controlled and uncontrolled regions throughout the crank cycle, along
with the switching states and dead points. Lightly shaded regions
are the controlled regions, and darkly shaded regions are the uncontrolled
regions.\label{fig:stim regions}}
\end{figure}

Cycling is achieved by switching between stimulation of the left and
right quadriceps muscle groups in an alternating pattern, where stimulation
of the quadriceps occurs outside of $\mathcal{Q}_{u}$ to avoid the
dead points and their neighborhoods where pedaling is inefficient.
Specifically, the switched control input $u^{s}$ is designed as
\begin{equation}
u^{s}\triangleq\begin{cases}
v & \mbox{if}\: q\in\mathcal{Q}_{c}\\
0 & \mbox{if}\: q\in\mathcal{Q}_{u}
\end{cases},\label{eq: switched control input}
\end{equation}
where $v\in\mathbb{R}$ is the stimulation control input. Substitution
of (\ref{eq: switched control input}) into (\ref{eq:Final General Dynamics Expression})
yields a switched system with autonomous state-dependent switching
as 
\begin{equation}
M\ddot{q}+V\dot{q}+G+\tau_{d}-\tau_{b}-P=\begin{cases}
B_{k}\Omega v & \mbox{if}\: q\in\mathcal{Q}_{c}\\
0 & \mbox{if}\: q\in\mathcal{Q}_{u}
\end{cases}.\label{eq: Switched System}
\end{equation}
Assuming that $q$ starts inside $\mathcal{Q}_{c}$, the known sequence
of switching states is defined as $\left\{ q_{n}^{on},\: q_{n}^{off}\right\} ,$
$n\in\left\{ 0,\:1,\:2,\:...\right\} $, where $q_{0}^{on}$ is the
initial crank angle and the switching states which follow are the
limit points of $\mathcal{Q}_{u}.$ The subsequent analysis is facilitated
by defining the corresponding sequence of switching times $\left\{ t_{n}^{on},\: t_{n}^{off}\right\} ,$
which are unknown a priori, where each on-time $t_{n}^{on}$ and off-time
$t_{n}^{off}$ occurs when $q$ reaches the corresponding on-angle
$q_{n}^{on}$ and off-angle $q_{n}^{off},$ respectively. Fig. \ref{fig:stim regions}
illustrates controlled and uncontrolled regions, along with the switching
states and dead points, throughout the crank cycle. 

\section{Control Development}

\subsection{Open-Loop Error System}

The control objective is to track a desired crank trajectory with
performance quantified by the tracking error signals $e_{1},\: e_{2}\in\mathbb{R}$,
defined as %
\begin{eqnarray}
e_{1} & \triangleq & q_{d}-q,\label{eq:e1}\\
e_{2} & \triangleq & \dot{e}_{1}+\alpha e_{1},\label{eq:e2}
\end{eqnarray}
where $q_{d}\in\mathbb{R}$ is the desired crank position, designed
so that its derivatives exist, and $\dot{q}_{d},\:\ddot{q}_{d}\in\mathcal{L}_{\infty},$
and $\alpha\in\mathbb{R}_{>0}$ is a\textbf{ }selectable constant.
Without loss of generality, $q_{d}$ is designed to monotonically
increase, i.e., stopping or backpedaling is not desired. Taking the
time derivative of (\ref{eq:e2}), multiplying by $M,$ and using
(\ref{eq: Switched System})-(\ref{eq:e2}) yields the following open-loop
error system:

\begin{equation}
M\dot{e}_{2}=\chi-Ve_{2}-\begin{cases}
B_{k}\Omega v & \mbox{if}\: q\in\mathcal{Q}_{c}\\
0 & \mbox{if}\: q\in\mathcal{Q}_{u}
\end{cases},\label{eq:OLES}
\end{equation}
where the auxiliary term $\chi\in\mathbb{R}$ is defined as

\begin{equation}
\chi\triangleq M\left(\ddot{q}_{d}+\alpha\dot{e}_{1}\right)+V\left(\dot{q}_{d}+\alpha e_{1}\right)+G+\tau_{d}-\tau_{b}-P.\label{eq: chi}
\end{equation}
Based on (\ref{eq: chi}) and Properties 1-6, $\chi$ can be bounded
as
\begin{equation}
\left|\chi\right|\leq c_{1}+c_{2}\left\Vert z\right\Vert +c_{3}\left\Vert z\right\Vert ^{2},\label{eq: chi bound}
\end{equation}
where $c_{1},\: c_{2},\: c_{3}\in\mathbb{R}_{>0}$ are known constants
and the error vector $z\in\mathbb{R}^{2}$ is defined as 
\begin{equation}
z\triangleq\left[\begin{array}{cc}
e_{1} & e_{2}\end{array}\right]^{T}.\label{eq: z}
\end{equation}

\subsection{Closed-Loop Error System}

Based on (\ref{eq:OLES}) and the subsequent stability analysis, the
control voltage input is designed as 
\begin{equation}
v\triangleq-k_{1}e_{2}-\left(k_{2}+k_{3}\left\Vert z\right\Vert +k_{4}\left\Vert z\right\Vert ^{2}\right)\mbox{sgn}\left(e_{2}\right),\label{eq:controller}
\end{equation}
where $\mbox{sgn}\left(\cdot\right)$ denotes the signum function
and $k_{1},\: k_{2},\: k_{3},\: k_{4}\in\mathbb{R}_{>0}$ are constant
control gains. After substituting (\ref{eq:controller}) into the
open-loop error system in (\ref{eq:OLES}), the following switched
closed-loop error system for $q\in\mathcal{Q}_{c}$ is obtained:
\begin{eqnarray}
M\dot{e}_{2} & = & \chi-Ve_{2}+B_{k}\Omega\biggl[k_{1}e_{2}\nonumber \\
 &  & +\biggl(k_{2}+k_{3}\left\Vert z\right\Vert +k_{4}\left\Vert z\right\Vert ^{2}\biggr)\mbox{sgn}\left(e_{2}\right)\biggr].\label{eq: CLES}
\end{eqnarray}
The controller in (\ref{eq:controller}) could also include $\left(B_{k}\right)^{-1}$
to cancel the preceding $B_{k}$ since it is known, resulting in less
conservative gain conditions. However, doing so would cause sharp
increases of the control input near the uncontrolled regions, depending
on the choice of $\varepsilon$, resulting in undesirably large magnitude
stimulation of the muscles in regions where their effectiveness is
low.

\section{Stability Analysis}

Let $V_{L}\::\:\mathbb{R}^{2}\rightarrow\mathbb{R}$ denote a continuously
differentiable, positive definite, radially unbounded, common Lyapunov-like
function defined as
\begin{equation}
V_{L}\triangleq\frac{1}{2}z^{T}Wz,\label{eq:VL}
\end{equation}
where the positive definite matrix $W\in\mathbb{R}^{2\times2}$ is
defined as 
\begin{equation}
W\triangleq\left[\begin{array}{cc}
1 & 0\\
0 & M
\end{array}\right].\label{eq: W}
\end{equation}
The function $V_{L}$ can be upper and lower bounded as
\begin{equation}
\lambda_{1}\left\Vert z\right\Vert ^{2}\leq V_{L}\leq\lambda_{2}\left\Vert z\right\Vert ^{2},\label{eq:VL bounds}
\end{equation}
where $\lambda_{1},\:\lambda_{2}\in\mathbb{R}_{>0}$ are known constants
defined as 
\[
\lambda_{1}\triangleq\mbox{min}\left(\frac{1}{2},\:\frac{c_{m}}{2}\right),\:\:\:\lambda_{2}\triangleq\mbox{max}\left(\frac{1}{2},\:\frac{c_{M}}{2}\right).
\]

\begin{thm}
\label{thm: decay bound}For $q\in\mathcal{Q}_{c}$, the closed-loop
error system in (\ref{eq: CLES}) is exponentially stable in the sense
that 
\begin{eqnarray}
\left\Vert z\left(t\right)\right\Vert  & \leq & \sqrt{\frac{\lambda_{2}}{\lambda_{1}}}\left\Vert z\left(t_{n}^{on}\right)\right\Vert e^{-\frac{\gamma_{1}}{2\lambda_{2}}\left(t-t_{n}^{on}\right)}\label{eq: z exp decay}
\end{eqnarray}
$\forall t\in\left(t_{n}^{on},\: t_{n}^{off}\right)$ and $\forall n$,
where $\gamma_{1}\in\mathbb{R}_{>0}$ is defined as
\begin{equation}
\gamma_{1}\triangleq\mathit{\mbox{min}}\left(\alpha-\frac{1}{2},\:\varepsilon c_{\Omega1}k_{1}-\frac{1}{2}\right),\label{eq: gamma1}
\end{equation}
provided the following gain conditions are satisfied:
\[
\alpha>\frac{1}{2},\: k_{1}>\frac{1}{2\varepsilon c_{\Omega1}},\: k_{2}\geq\frac{c_{1}}{\varepsilon c_{\Omega1}},
\]
\begin{equation}
k_{3}\geq\frac{c_{2}}{\varepsilon c_{\Omega1}},\: k_{4}\geq\frac{c_{3}}{\varepsilon c_{\Omega1}}.\label{eq: gains 1}
\end{equation}
\end{thm}
\begin{IEEEproof}
Let $z\left(t\right)$ for $t\in\left(t_{n}^{on},\: t_{n}^{off}\right)$
be a Filippov solution to the differential inclusion $\dot{z}\in K\left[h\right]\left(z\right),$
where $K\left[\cdot\right]$ is defined as in \cite{Filippov1964}
and where $h\::\:\mathbb{R}\times\mathbb{R}\rightarrow\mathbb{R}^{2}$
is defined using (\ref{eq:e2}) and (\ref{eq: CLES}) as
\begin{equation}
h\triangleq\left[\begin{array}{c}
e_{2}-\alpha e_{1}\\
M^{-1}\biggl\{\chi-Ve_{2}+B_{k}\Omega\biggl[k_{1}e_{2}\:\:\:\:\:\:\:\:\:\:\:\:\:\:\:\\
\:\:\:+\left(k_{2}+k_{3}\left\Vert z\right\Vert +k_{4}\left\Vert z\right\Vert ^{2}\right)\mbox{sgn}\left(e_{2}\right)\biggr]\biggr\}
\end{array}\right].\label{eq: h}
\end{equation}
The time derivative of (\ref{eq:VL}) exists almost everywhere (a.e.),
i.e., for almost all $t\in\left(t_{n}^{on},\: t_{n}^{off}\right),$
and $\dot{V}_{L}\left(z\right)\overset{a.e.}{\in}\dot{\tilde{V}}_{L}\left(z\right),$
where $\dot{\tilde{V}}_{L}$ is the generalized time derivative of
(\ref{eq:VL}) along the Filippov trajectories of $\dot{z}=h\left(z\right)$
and is defined as \cite{Paden1987}
\[
\dot{\tilde{V}}_{L}\triangleq\underset{\xi\in\partial V_{L}\left(z\right)}{\cap}\xi^{T}K\left[\begin{array}{c}
h\left(z\right)\\
1
\end{array}\right],
\]
where $\partial V_{L}$ is the generalized gradient of $V_{L}$. Since
$V_{L}$ is continuously differentiable in $z$, $\partial V_{L}=\left\{ \nabla V_{L}\right\} ;$
thus, 
\[
\dot{\tilde{V}}_{L}\subseteq\left[\begin{array}{c}
z^{T}W\\
\frac{1}{2}z^{T}\dot{W}z
\end{array}\right]^{T}K\left[\begin{array}{c}
h\left(z\right)\\
1
\end{array}\right].
\]
Using the calculus of $K$ from \cite{Paden1987}, substituting (\ref{eq: h}),
and using (\ref{eq: W}) to simplify the resulting expression yields
\begin{eqnarray}
\dot{\tilde{V}}_{L} & \subseteq & -\alpha e_{1}^{2}+e_{1}e_{2}+\chi e_{2}+\left(\frac{1}{2}\dot{M}-V\right)e_{2}^{2}\nonumber \\
 &  & +B_{k}\Omega\left(k_{1}e_{2}^{2}\right)+B_{k}\Omega\biggl(k_{2}+k_{3}\left\Vert z\right\Vert \nonumber \\
 &  & +k_{4}\left\Vert z\right\Vert ^{2}\biggr)K\left[\mbox{sgn}\right]\left(e_{2}\right)e_{2},\label{eq: V_dot}
\end{eqnarray}
where $K\left[\mbox{sgn}\right]\left(e_{2}\right)=1$ $\mbox{if}\: e_{2}>0,\:\left[-1,\:1\right]\:\mbox{if}\: e_{2}=0,$
and $-1\:\mbox{if}\: e_{2}<0.$ Using Property 8 and the fact that
$\dot{V}_{L}\left(z\right)\overset{a.e.}{\in}\dot{\tilde{V}}_{L}\left(z\right)$
allows (\ref{eq: V_dot}) to be rewritten as
\begin{eqnarray}
\dot{V}_{L} & \overset{a.e.}{=} & -\alpha e_{1}^{2}+e_{1}e_{2}+\chi e_{2}+B_{k}\Omega\left(k_{1}e_{2}^{2}\right)\nonumber \\
 &  & +B_{k}\Omega\left(k_{2}+k_{3}\left\Vert z\right\Vert +k_{4}\left\Vert z\right\Vert ^{2}\right)\left|e_{2}\right|.\label{eq: V_dot 1-1}
\end{eqnarray}
Note that $K\left[\mbox{sgn}\right]\left(e_{2}\right)$ is only set-valued
for $e_{2}=0,$ so that $K\left[\mbox{sgn}\right]\left(e_{2}\right)e_{2}$
may be rewritten simply as $\left|e_{2}\right|,$ as was done in (\ref{eq: V_dot 1-1}).
By using Young's inequality, (\ref{eq: chi bound}), Property 7, and
the fact that $B_{k}<-\varepsilon$ for $q\in\mathcal{Q}_{c}$, (\ref{eq: V_dot 1-1})
can be upper bounded as
\begin{eqnarray}
\dot{V}_{L} & \overset{a.e.}{\leq} & -\left(\alpha-\frac{1}{2}\right)e_{1}^{2}-\left(\varepsilon c_{\Omega1}k_{1}-\frac{1}{2}\right)e_{2}^{2}\nonumber \\
 &  & +\left(c_{1}-\varepsilon c_{\Omega1}k_{2}\right)\left|e_{2}\right|+\left(c_{2}-\varepsilon c_{\Omega1}k_{3}\right)\left\Vert z\right\Vert \left|e_{2}\right|\nonumber \\
 &  & +\left(c_{3}-\varepsilon c_{\Omega1}k_{4}\right)\left\Vert z\right\Vert ^{2}\left|e_{2}\right|.\label{eq: V_dot 2}
\end{eqnarray}
Provided the gain conditions in (\ref{eq: gains 1}) are satisfied,
(\ref{eq:VL bounds}) can be used to rewrite (\ref{eq: V_dot 2})
as 
\begin{equation}
\dot{V}_{L}\overset{a.e.}{\leq}-\frac{\gamma_{1}}{\lambda_{2}}V_{L},\label{eq: V_dot 3}
\end{equation}
where $\gamma_{1}$ was defined in (\ref{eq: gamma1}). The inequality
in (\ref{eq: V_dot 3}) can be rewritten as
\[
e^{\frac{\gamma_{1}}{\lambda_{2}}\left(t-t_{n}^{on}\right)}\left(\dot{V}_{L}+\frac{\gamma_{1}}{\lambda_{2}}V_{L}\right)\overset{a.e.}{\leq}0,
\]
which is equivalent to the following expression:
\begin{equation}
\frac{d}{dt}\left(V_{L}e^{\frac{\gamma_{1}}{\lambda_{2}}\left(t-t_{n}^{on}\right)}\right)\overset{a.e.}{\leq}0.\label{eq: V_dot 4}
\end{equation}
Taking the Lebesgue integral of (\ref{eq: V_dot 4}) and recognizing
that the integrand on the left-hand side is absolutely continuous
allows the Fundamental Theorem of Calculus to be used to yield
\[
V_{L}e^{\frac{\gamma_{1}}{\lambda_{2}}\left(t-t_{n}^{on}\right)}\leq C,
\]
where $C\in\mathbb{R}$ is a constant of integration equal to $V_{L}\left(z\left(t_{n}^{on}\right)\right)$.
Therefore, 
\begin{eqnarray}
V_{L}\left(z\left(t\right)\right) & \leq & V_{L}\left(z\left(t_{n}^{on}\right)\right)e^{-\frac{\gamma_{1}}{\lambda_{2}}\left(t-t_{n}^{on}\right)}\nonumber \\
 &  & \:\:\:\forall t\in\left(t_{n}^{on},\: t_{n}^{off}\right),\:\:\:\forall n.\label{eq: V decay}
\end{eqnarray}
Using (\ref{eq:VL bounds}) to rewrite (\ref{eq: V decay}) and performing
some algebraic manipulation yields (\ref{eq: z exp decay}).\end{IEEEproof}
\begin{rem}
Theorem \ref{thm: decay bound} guarantees that desired crank trajectories
can be tracked with exponential convergence, provided that the crank
angle does not exit the controlled region. Thus, if the controlled
regions and desired trajectories are designed appropriately, the controller
in (\ref{eq:controller}) yields exponential tracking of the desired
trajectories for all time. However, if the crank position exits the
controlled region, the system becomes uncontrolled and the following
theorem details the resulting error system behavior. \end{rem}
\begin{thm}
\label{thm: z growth}For $q\in\mathcal{Q}_{u}$, the closed-loop
error system in (\ref{eq: CLES}) can be upper bounded as
\begin{eqnarray}
\left\Vert z\left(t\right)\right\Vert  & \leq & \frac{1}{2a_{1}\sqrt{\lambda_{1}}}\Biggl\{ a_{3}\mbox{tan}\Biggl[\frac{a_{3}}{4}\left(t-t_{n}^{off}\right)\nonumber \\
 &  & +\mbox{tan}^{-1}\left(\frac{2a_{1}}{a_{3}}\sqrt{\lambda_{2}}\left\Vert z\left(t_{n}^{off}\right)\right\Vert +\frac{a_{2}}{a_{3}}\right)\Biggr]-a_{2}\Biggr\}\nonumber \\
\label{eq: z growth}
\end{eqnarray}
$\forall t\in\left[t_{n}^{off},\: t_{n+1}^{on}\right]$ and $\forall n,$
provided the time spent in the uncontrolled region $\Delta t_{n}^{off}\triangleq t_{n+1}^{on}-t_{n}^{off}$
is sufficiently small, in the sense that
\begin{eqnarray}
\Delta t_{n}^{off} & < & \frac{1}{a_{3}}\biggl[2\pi-4\mbox{tan}^{-1}\biggl(\frac{2a_{1}}{a_{3}}\sqrt{\lambda_{2}}\left\Vert z\left(t_{n}^{off}\right)\right\Vert \nonumber \\
 &  & \:\:\:\:\:\:\:\:\:\:\:\:\:\:\:\:+\frac{a_{2}}{a_{3}}\biggr)\Biggr]\label{eq: tangent RDT-1}
\end{eqnarray}
$\forall n,$ where $a_{1},\: a_{2},\: a_{3}\in\mathbb{R}_{>0}$ are
known constants defined as
\begin{equation}
a_{1}\triangleq\frac{c_{3}}{\left(\lambda_{1}\right)^{\frac{3}{2}}},\:\:\: a_{2}\triangleq\frac{c_{2}+\frac{1}{2}}{\lambda_{1}},\:\:\: a_{3}\triangleq\frac{4a_{1}c_{1}}{\sqrt{\lambda_{1}}}-a_{2}^{2}.\label{eq: ABC}
\end{equation}
\end{thm}
\begin{IEEEproof}
The time derivative of (\ref{eq:VL}) for all $t\in\left[t_{n}^{off},\: t_{n+1}^{on}\right]$
can be expressed using (\ref{eq:e2}), (\ref{eq: CLES}), and Property
8 as
\begin{equation}
\dot{V}_{L}=-\alpha e_{1}^{2}+e_{1}e_{2}+\chi e_{2}.\label{eq: V_dot off}
\end{equation}
Young's inequality, (\ref{eq: chi bound}), and (\ref{eq: z}) allow
(\ref{eq: V_dot off}) to be upper bounded as 
\begin{equation}
\dot{V}_{L}\leq c_{1}\left\Vert z\right\Vert +\left(c_{2}+\frac{1}{2}\right)\left\Vert z\right\Vert ^{2}+c_{3}\left\Vert z\right\Vert ^{3}.\label{eq: V_dot off bound 1}
\end{equation}
Using (\ref{eq:VL bounds}), (\ref{eq: V_dot off bound 1}) can be
upper bounded as
\begin{equation}
\dot{V}_{L}\leq\frac{c_{1}}{\sqrt{\lambda_{1}}}V_{L}^{\frac{1}{2}}+\frac{c_{2}+\frac{1}{2}}{\lambda_{1}}V_{L}+\frac{c_{3}}{\left(\lambda_{1}\right)^{\frac{3}{2}}}V_{L}^{\frac{3}{2}}.\label{eq: V_dot off bound 2}
\end{equation}
The solution to (\ref{eq: V_dot off bound 2}) yields the following
upper bound on $V_{L}$ in the uncontrolled region:
\begin{eqnarray}
V_{L}\left(z\left(t\right)\right) & \leq & \frac{1}{4a_{1}^{2}}\Biggl\{ a_{3}\mbox{tan}\Biggl[\frac{a_{3}}{4}\left(t-t_{n}^{off}\right)\nonumber \\
 &  & +\mbox{tan}^{-1}\left(\frac{2a_{1}}{a_{3}}\sqrt{V_{L}\left(z\left(t_{n}^{off}\right)\right)}+\frac{a_{2}}{a_{3}}\right)\Biggr]\nonumber \\
 &  & \:\:\:\:\:\:\:\:\:\:\:\:\:\:\:\:\:\:\:\:-a_{2}\Biggr\}^{2}\label{eq: V growth}
\end{eqnarray}
$\forall t\in\left[t_{n}^{off},\: t_{n+1}^{on}\right]$ and $\forall n,$
where $a_{1},\: a_{2},$ and $a_{3}$ were defined in (\ref{eq: ABC}).
Using (\ref{eq:VL bounds}) to rewrite (\ref{eq: V growth}) and performing
some algebraic manipulation yields (\ref{eq: z growth}). \end{IEEEproof}
\begin{rem}
The bound in (\ref{eq: z growth}) has a finite escape time, so $\left\Vert z\right\Vert $
may become unbounded unless the reverse dwell-time (RDT) condition
in (\ref{eq: tangent RDT-1}) is satisfied. In other words, the argument
of $\mbox{tan}\left(\cdot\right)$ in (\ref{eq: z growth}) must be
less than $\frac{\pi}{2}$ to ensure boundedness of $\left\Vert z\right\Vert $.
The following propositions and subsequent proofs detail how the RDT
condition may be satisfied.\end{rem}
\begin{prop}
\label{delta t Proposition }The time spent in the $n^{th}$ controlled
region $\Delta t_{n}^{on}\triangleq t_{n}^{off}-t_{n}^{on}$ has a
known positive lower bound $\Delta t_{min}^{on}\in\mathbb{R}_{>0}$
such that 
\[
\underset{n}{\mbox{min}}\:\Delta t_{n}^{on}\geq\Delta t_{min}^{on}>0.
\]
\end{prop}
\begin{IEEEproof}
The time spent in the $n^{th}$ controlled region $\Delta t_{n}^{on}$
can be described using the Mean Value Theorem as
\begin{equation}
\Delta t_{n}^{on}=\frac{\Delta q_{n}^{on}}{\dot{q}\left(\xi_{n}^{on}\right)},\label{eq: delta t on}
\end{equation}
where $\Delta q_{n}^{on}\triangleq q\left(t_{n}^{off}\right)-q\left(t_{n}^{on}\right)$
is the length of the $n^{th}$ controlled region, which is constant
for all $n\geq1$ and is smallest for $n=0$ in this development,
and $\dot{q}\left(\xi_{n}^{on}\right)\in\mathbb{R}$ is the average
crank velocity through the $n^{th}$ controlled region. Using (\ref{eq:e1})
and (\ref{eq:e2}), $\dot{q}$ can be upper bounded as
\begin{equation}
\dot{q}\leq\dot{q}_{d}+\left(1+\alpha\right)\left\Vert z\right\Vert .\label{eq: q_dot bound}
\end{equation}
Then, using the fact that $V_{L}$ monotonically decreases in the
controlled regions together with (\ref{eq:VL bounds}) and (\ref{eq: q_dot bound})
allows the average crank velocity $\dot{q}\left(\xi_{n}^{on}\right)$
to be upper bounded as
\begin{equation}
\dot{q}\left(\xi_{n}^{on}\right)\leq\underset{\forall t}{\mbox{max}}\:\dot{q}_{d}+\left(1+\alpha\right)\sqrt{\frac{\lambda_{2}}{\lambda_{1}}}\left\Vert z\left(t_{n}^{on}\right)\right\Vert .\label{eq: average q_dot on bound}
\end{equation}
Therefore, (\ref{eq: delta t on}) can be lower bounded using (\ref{eq: average q_dot on bound})
as
\begin{equation}
\Delta t_{n}^{on}\geq\frac{\Delta q_{n}^{on}}{\underset{\forall t}{\mbox{max}}\:\dot{q}_{d}+\left(1+\alpha\right)\sqrt{\frac{\lambda_{2}}{\lambda_{1}}}\left\Vert z\left(t_{n}^{on}\right)\right\Vert }.\label{eq: ton lower bound}
\end{equation}
For a given $\Delta t_{min}^{on}$, (\ref{eq: ton lower bound}) can
be used to determine that
\[
\frac{\Delta q_{n}^{on}}{\underset{\forall t}{\mbox{max}}\:\dot{q}_{d}+\left(1+\alpha\right)\sqrt{\frac{\lambda_{2}}{\lambda_{1}}}\left\Vert z\left(t_{n}^{on}\right)\right\Vert }\geq\Delta t_{min}^{on},
\]
which can be satisfied by selecting the desired trajectory as
\begin{equation}
\underset{\forall t}{\mbox{max}}\:\dot{q}_{d}\leq\frac{\Delta q_{n}^{on}}{\Delta t_{min}^{on}}-\left(1+\alpha\right)\sqrt{\frac{\lambda_{2}}{\lambda_{1}}}\left\Vert z\left(t_{n}^{on}\right)\right\Vert .\label{eq: max qd_dot}
\end{equation}
The inequality in (\ref{eq: max qd_dot}) provides an upper bound
on the desired velocity which guarantees that $\Delta t_{n}^{on}$
is greater than a given minimum $\Delta t_{min}^{on}$. \end{IEEEproof}
\begin{assumption}
\label{ass: q_crit}Since $\Delta t_{n}^{off}\propto\dot{q}\left(t_{n}^{off}\right)^{-1}$,
there exists a known initial velocity $\dot{q}\left(t_{n}^{off}\right),$
denoted as the critical velocity $\dot{q}^{crit}\in\mathbb{R}_{>0}$,
which satisfies (\ref{eq: tangent RDT-1}) for all $n$. This assumption
is mild in the sense that, given a desired $\Delta t_{n}^{off},$
the critical velocity can be experimentally determined for an individual
system configuration or numerically calculated for a wide range of
individual or cycle configurations.\textbf{}%
\end{assumption}
\begin{prop}
The time spent in the $n^{th}$ uncontrolled region $\Delta t_{n}^{off}$
has a known positive upper bound $\Delta t_{max}^{off}\in\mathbb{R}_{>0}$
that satisfies 
\begin{eqnarray}
\Delta t_{max}^{off} & < & \frac{1}{a_{3}}\Biggl[2\pi-4\mbox{tan}^{-1}\biggl(\frac{2a_{1}}{a_{3}}\sqrt{\lambda_{2}}\left\Vert z\left(t_{n}^{off}\right)\right\Vert \nonumber \\
 &  & \:\:\:\:\:\:\:\:\:\:\:\:\:\:\:\:+\frac{a_{2}}{a_{3}}\biggr)\Biggr]\:\:\:\forall n.\label{eq: max tangent RDT}
\end{eqnarray}
\end{prop}
\begin{IEEEproof}
The crank's entrance into the uncontrolled region can be likened to
a ballistic event, where the crank is positioned at $q\left(t_{n}^{off}\right)$
and released with initial velocity $\dot{q}\left(t_{n}^{off}\right).$
Therefore, specifying a desired $\Delta t_{max}^{off}$ is equivalent
to requiring the crank to ballistically (i.e., only under the influence
of passive dynamics) traverse the length of the uncontrolled region,
$\Delta q_{n}^{off}$, in a sufficiently short amount of time. The
only controllable factors affecting the behavior of the crank in the
uncontrolled region are the initial conditions. Since $q\left(t_{n}^{off}\right)$
is predetermined by selection of $\varepsilon$, then only the initial
velocity $\dot{q}\left(t_{n}^{off}\right)$ can be used to guarantee
that the total time spent in the $n^{th}$ uncontrolled region is
less than $\Delta t_{max}^{off}.$ That is, using Assumption \ref{ass: q_crit},
it can be demonstrated that (\ref{eq: tangent RDT-1}) is satisfied
provided  
\begin{equation}
\dot{q}\left(t_{n}^{off}\right)\geq\dot{q}^{crit}.\label{eq: qcrit result}
\end{equation}
Using (\ref{eq:e1}) and (\ref{eq:e2}), $\dot{q}\left(t_{n}^{off}\right)$
can be lower bounded as
\begin{equation}
\dot{q}\left(t_{n}^{off}\right)\geq\dot{q}_{d}\left(t_{n}^{off}\right)-\left(1+\alpha\right)\left\Vert z\left(t_{n}^{off}\right)\right\Vert .\label{eq: q_dot lower bound}
\end{equation}
Combining (\ref{eq: qcrit result}) and (\ref{eq: q_dot lower bound}),
the following sufficient condition for the desired crank velocity
at the $n^{th}$ off-time which guarantees (\ref{eq: qcrit result})
can be developed:
\begin{equation}
\dot{q}_{d}\left(t_{n}^{off}\right)\geq\dot{q}^{crit}+\left(1+\alpha\right)\left\Vert z\left(t_{n}^{off}\right)\right\Vert .\label{eq: min qd_dot}
\end{equation}
Furthermore, (\ref{eq: z exp decay}) can be used to obtain a sufficient
condition for (\ref{eq: min qd_dot}) in terms of the initial conditions
of each cycle as 
\begin{equation}
\dot{q}_{d}\left(t_{n}^{off}\right)\geq\dot{q}^{crit}+\left(1+\alpha\right)\sqrt{\frac{\lambda_{2}}{\lambda_{1}}}\left\Vert z\left(t_{n}^{on}\right)\right\Vert e^{-\frac{\gamma_{1}}{2\lambda_{2}}\Delta t_{min}^{on}}.\label{eq: final min qd_dot}
\end{equation}
\end{IEEEproof}
\begin{thm}
\label{thm: UB Theorem}The closed-loop error system in (\ref{eq: CLES})
is ultimately bounded in the sense that, as the number of crank cycles
approaches infinity (i.e., as $n\rightarrow\infty$), $\left\Vert z\left(t\right)\right\Vert $
converges to a ball with constant radius $d\in\mathbb{R}_{>0}$. %
\end{thm}
\begin{IEEEproof}
There are three possible scenarios which describe the behavior of
$V_{L}$ with each cycle: (1) $V_{L}\left(z\left(t_{n+1}^{on}\right)\right)<V_{L}\left(z\left(t_{n}^{on}\right)\right),$
(2) $V_{L}\left(z\left(t_{n+1}^{on}\right)\right)>V_{L}\left(z\left(t_{n}^{on}\right)\right),$
and (3) $V_{L}\left(z\left(t_{n+1}^{on}\right)\right)=V_{L}\left(z\left(t_{n}^{on}\right)\right).$
The potential decay and growth of $V_{L}$ in the controlled and uncontrolled
regions, respectively, dictates which of these three behaviors will
occur. In scenario (1), the decay is greater than the growth, causing
$V_{L}$ to decrease with each cycle. Conversely, in scenario (2),
the growth is greater than the decay, causing $V_{L}$ to grow with
each cycle. Since the amount of decay or growth is proportional to
the initial conditions for each region, $V_{L}\left(z\left(t_{n}^{on}\right)\right)$
and $V_{L}\left(z\left(t_{n}^{off}\right)\right)$, eventually (as
$n\rightarrow\infty$) the magnitude of the potential decay will equal
the magnitude of the potential growth, resulting in scenario (3) and
an ultimate bound on $V_{L}$. 

Suppose $V_{L}\left(z\left(t_{n}^{on}\right)\right)$ reaches the
ultimate bound $\overline{d}$ after $N$ cycles. Then, $V_{L}\left(z\left(t_{n+1}^{on}\right)\right)=\overline{d}$
$\forall n\geq N-1$. Then $\overline{d}$ can be found by considering
the most conservative case where the minimum possible decay and the
maximum possible growth are equal, i.e., by considering $V_{L}\left(z\left(t_{n+1}^{on}\right)\right)=\overline{d}$
with $\Delta t_{n}^{on}=\Delta t_{min}^{on}$ and $\Delta t_{n}^{off}=\Delta t_{max}^{off}$
for all $n.$ Therefore, the most conservative ultimate bound on $V_{L}$
can be found by solving the following equation for $\overline{d}$
using (\ref{eq: V decay}) with $\Delta t_{min}^{on}$ and (\ref{eq: V growth})
with $\Delta t_{max}^{off}:$
\begin{eqnarray}
\overline{d} & = & \frac{1}{4a_{1}^{2}}\Biggl\{ a_{3}\mbox{tan}\Biggl[\frac{a_{3}}{4}\Delta t_{max}^{off}\nonumber \\
 &  & \:\:\:\:\:+\mbox{tan}^{-1}\left(\frac{2a_{1}}{a_{3}}\sqrt{\overline{d}}e^{-\frac{\gamma_{1}}{2\lambda_{2}}\Delta t_{min}^{on}}+\frac{a_{2}}{a_{3}}\right)\Biggr]-a_{2}\Biggr\}^{2}.\nonumber \\
\label{eq: bound equation}
\end{eqnarray}
Algebraic manipulation of (\ref{eq: bound equation}) gives a quadratic
polynomial in $\sqrt{\overline{d}}$ in the following form:
\begin{equation}
b_{1}\overline{d}+b_{2}\sqrt{\overline{d}}+b_{3}=0,\label{eq: polynomial ultimate bound}
\end{equation}
where $b_{1},\: b_{2},\: b_{3}\in\mathbb{R}$ are known constants
defined as
\begin{eqnarray*}
b_{1} & \triangleq & -4a_{1}^{2}\mbox{tan}\left(\frac{a_{3}}{4}\Delta t_{max}^{off}\right)e^{-\frac{\gamma_{1}}{2\lambda_{2}}\Delta t_{min}^{on}},\\
b_{2} & \triangleq & 2a_{1}a_{3}\left(1-e^{-\frac{\gamma_{1}}{2\lambda_{2}}\Delta t_{min}^{on}}\right)\\
 &  & -2a_{1}a_{2}\mbox{tan}\left(\frac{a_{3}}{4}\Delta t_{max}^{off}\right)\left(1+e^{-\frac{\gamma_{1}}{2\lambda_{2}}\Delta t_{min}^{on}}\right),\\
b_{3} & \triangleq & -\left(a_{2}^{2}+a_{3}^{2}\right)\mbox{tan}\left(\frac{a_{3}}{4}\Delta t_{max}^{off}\right).
\end{eqnarray*}
Solving (\ref{eq: polynomial ultimate bound}) for $\overline{d}$,
provided $b_{1},\: b_{2},\: b_{3}\neq0,$ gives the resulting ultimate
bound on $V_{L}\left(z\left(t_{n}^{on}\right)\right)\:\forall n\geq N$:
\[
\overline{d}=\left(\frac{-b_{2}+\sqrt{b_{2}^{2}-4b_{1}b_{3}}}{2b_{1}}\right)^{2}.
\]
Additionally, the ultimate bound on $V_{L}\left(z\left(t_{n}^{off}\right)\right)\:\forall n\geq N$
can be found by considering the minimum decay of $V_{L}$ in the controlled
regions after the ultimate bound has been reached. This lower bound,
denoted by $\underline{d}\in\mathbb{R}_{\geq0}$, is $\underline{d}\triangleq\overline{d}e^{-\frac{\gamma_{1}}{\lambda_{2}}\Delta t_{min}^{on}}.$
Since the bounds on $V_{L}$ in the controlled and uncontrolled regions
strictly decrease and increase, respectively, $V_{L}\left(z\left(t\right)\right)\leq\overline{d}\:\forall t\geq t_{N}^{on}$
when $V_{L}\left(z\left(t_{N}^{on}\right)\right)\leq\overline{d},$
or, equivalently, $\forall t\geq t_{N}^{off}$ when $V_{L}\left(z\left(t_{N}^{off}\right)\right)\leq\underline{d}.$
In other words, if the magnitude of $V_{L}$ is smaller than $\overline{d}$
when the controller is switched on or smaller than $\underline{d}$
when the controller is switched off, then $V_{L}$ will henceforth
remain smaller than $\overline{d}$. From (\ref{eq:VL bounds}), $\left\Vert z\left(t\right)\right\Vert $
converges to a ball with constant radius, i.e., $\left\Vert z\left(t\right)\right\Vert \rightarrow d\:\:\:\mbox{as}\:\:\: n\rightarrow\infty,$
where $d\in\mathbb{R}_{>0}$ is a constant defined as $d\triangleq\sqrt{\overline{d}/\lambda_{1}}.$%

\end{IEEEproof}

\section{Experimental Results}

An FES-cycling experiment was performed on an able-bodied male subject
age 24, height 186 cm, and weight 78 kg, with written informed consent
approved by the University of Florida Institutional Review Board.
The goal of this experiment was to demonstrate the tracking performance
and robustness of the controller in (\ref{eq:controller}). The experiment
was ended if $90$ revolutions had been completed, the control input
(pulse width) saturated at $400$ microseconds, or the subject reported
significant discomfort. An able-bodied subject was recruited for this
experiment as the response of nonimpaired subjects to electrical stimulation
has been reported as similar to the response of paraplegic subjects
\cite{Chang1997,Kurosawa2005,Jezernik2004,Hausdorff1991}. During
the experiment, the subject was instructed to relax and was given
no indication of the control performance. 

A fixed-gear stationary recumbent cycle was equipped with an optical
encoder to measure the crank angle and custom pedals upon which high-topped
orthotic walking boots were affixed. The purpose of the boots was
to fix the rider's feet to the pedals, hold the ankle position at
90 deg, and maintain sagittal alignment of the legs. The cycle has
an adjustable seat and a magnetically braked flywheel with sixteen
levels of resistance. The cycle seat position was adjusted for the
comfort of the rider, provided that the subject's knees could not
hyperextend while cycling. Geometric parameters of the stationary
cycle and subject were measured prior to the experiment. The following
distances were measured and used to calculate $B_{k}$ for the subject:
greater trochanter to lateral femoral condyle (thigh length), lateral
femoral condyle to pedal axis (effective shank length), and the horizontal
and vertical distance from the greater trochanter to the cycle crank
axis (seat position). The distance between the pedal axis and the
cycle crank axis (pedal length) was also measured. Electrodes were
then placed on the anterior distal-medial and proximal-lateral portions
of the subject's left and right thighs. 

A current-controlled stimulator (RehaStim, Hasomed, GmbH, Germany)
was used to stimulate the subject's quadriceps femoris muscle groups
through bipolar self-adhesive $3"\times5"$ PALS$^{\lyxmathsym{\textregistered}}$
Platinum oval electrodes (provided by Axelgaard Manufacturing Co.).
A personal computer equipped with data acquisition hardware and software
was used to read the encoder signal, calculate the control input,
and command the stimulator. Stimulation was conducted at a frequency
of $40$ Hz with a constant amplitude of $100$ mA and a variable
pulsewidth dictated by the controller in (\ref{eq:controller}).

The desired crank position and velocity were given in radians and
radians per second, respectively, as 
\begin{eqnarray}
q_{d} & \triangleq & 3.665\left(t-t_{0}^{on}\right)-\dot{q}_{d}+q_{0}^{on},\label{eq: qd}\\
\dot{q}_{d} & \triangleq & 3.665\left[1-\mbox{exp}\left(t_{0}^{on}-t\right)\right].\label{eq: qd_dot}
\end{eqnarray}
The trajectories in (\ref{eq: qd}) and (\ref{eq: qd_dot}) ensured
that the desired velocity started at 0 rpm and exponentially approached
35 rpm.%
{} The following control gains were found to be effective in preliminary
testing and were used for the experiment:
\[
\alpha=7,\: k_{1}=10,\: k_{2}=0.1,\: k_{3}=0.1,\: k_{4}=0.1.
\]
The stimulation region was determined by defining $\varepsilon\triangleq\frac{1}{2}\mbox{max}\left(\left|B_{k}^{s}\right|\right)=0.2739$
for the subject.

\begin{figure}
\begin{centering}
\includegraphics[width=1\columnwidth]{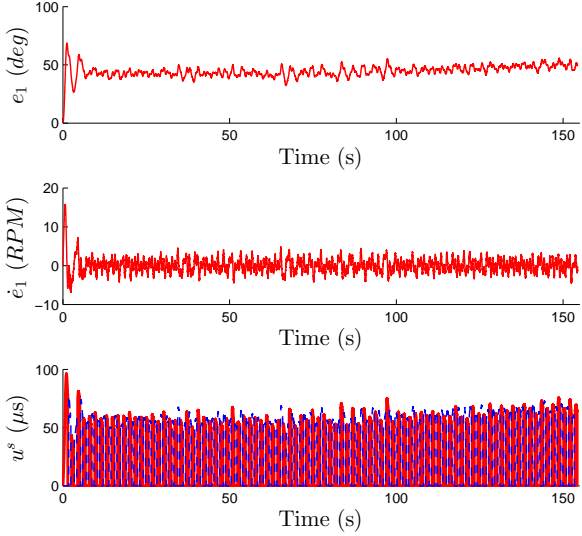}\caption{Tracking performance and input of the switched FES-cycling controller.\label{fig:Tracking-performance}}

\par\end{centering}

\end{figure}

\begin{figure}
\begin{centering}
\includegraphics[width=1\columnwidth]{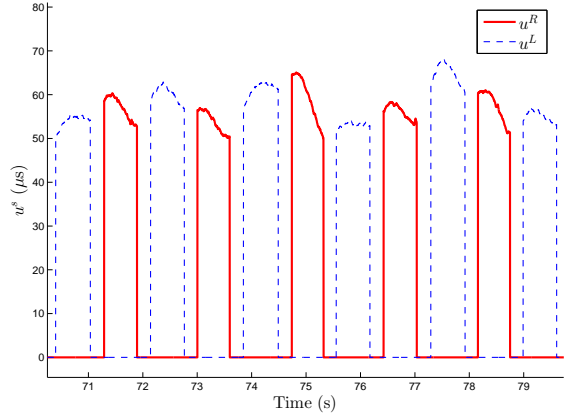}\caption{Switched control input $u^{s}$ over a subset of the total experiment
runtime illustrating the switching behavior of the controller.\label{fig:Switched-control-input}}

\par\end{centering}

\end{figure}

Fig. \ref{fig:Tracking-performance} depicts the resulting crank position
tracking error $e_{1}$, cadence tracking error $\dot{e}_{1},$ and
switched control input $u^{s}$ from the experiment. 
Note that the error is ultimately bounded and that the control input
switches between muscle groups without overlap according to the stimulation
regions defined in (\ref{eq: controlled regions}). Fig. \ref{fig:Switched-control-input}
illustrates the switching behavior of the control input in (\ref{eq: switched control input})

\section{Conclusion}

An FES-cycling stimulation pattern for the human quadriceps femoris
muscle groups is derived from torque transfer ratios which define
the effectiveness of knee torques at producing positive torque about
a stationary cycle's crank. The design of the stimulation pattern
allows for arbitrary resizing of the stimulation regions through choice
of the constant $\varepsilon$, provided they have nonzero length
and do not contain the cycling dead points. The results of \cite{Idso.Johansen.ea2004}
suggest that $\varepsilon$ should be made as large as possible to
minimize the stimulation region while maximizing stimulation intensity
for optimal metabolic efficiency. 

Since the controller switches between muscle groups based on the state-dependent
stimulation pattern and there exist regions where no control is applied,
the resulting system is a switched system with autonomous state-dependent
switching which must satisfy a reverse dwell-time condition (RDT).
The nonlinear nature of the system along with parametric uncertainty
and the presence of an unknown bounded disturbance make the RDT condition
uncertain, so a switched sliding-mode controller is designed and Lyapunov
methods for switched systems are utilized to develop sufficient conditions
on the control gains and the desired trajectory which ensure the RDT
condition is satisfied. Specifically, if the desired trajectory is
designed to satisfy (\ref{eq: max qd_dot}) and (\ref{eq: final min qd_dot}),
then (\ref{eq: tangent RDT-1}) is satisfied and the result of Theorem
\ref{thm: UB Theorem} follows. Note that (\ref{eq: final min qd_dot})
only restricts the desired trajectory at the off-times, giving freedom
to design the desired trajectory everywhere else, provided $q_{d}$
and $\dot{q}_{d}$ are continuously differentiable and that the maximum
value of $\dot{q}_{d}$ satisfies (\ref{eq: max qd_dot}).

The switched controller guarantees ultimately bounded tracking in
the sense that the norm of the error vector, $\left\Vert z\left(t\right)\right\Vert ,$
converges to a ball of constant radius $d,$ and experimental results
are provided which demonstrate the performance of the controller.
The size of the ultimate bound depends on $\varepsilon$, the control
gains, and the system parameters. The stimulation pattern and switched
controller may enable improved efficiency and power output in FES-cycling
systems. Future work will focus on further experimental verification
of this control strategy, its effect on efficiency and power output,
and stimulation of additional muscle groups.

\bibliographystyle{IEEEtran}
\bibliography{ncr,master}

\begin{thebibliography}{10}
\providecommand{\url}[1]{#1}
\csname url@samestyle\endcsname
\providecommand{\newblock}{\relax}
\providecommand{\bibinfo}[2]{#2}
\providecommand{\BIBentrySTDinterwordspacing}{\spaceskip=0pt\relax}
\providecommand{\BIBentryALTinterwordstretchfactor}{4}
\providecommand{\BIBentryALTinterwordspacing}{\spaceskip=\fontdimen2\font plus
\BIBentryALTinterwordstretchfactor\fontdimen3\font minus
  \fontdimen4\font\relax}
\providecommand{\BIBforeignlanguage}[2]{{%
\expandafter\ifx\csname l@#1\endcsname\relax
\typeout{** WARNING: IEEEtran.bst: No hyphenation pattern has been}%
\typeout{** loaded for the language `#1'. Using the pattern for}%
\typeout{** the default language instead.}%
\else
\language=\csname l@#1\endcsname
\fi
#2}}
\providecommand{\BIBdecl}{\relax}
\BIBdecl

\bibitem{Phillips.Petrofsky.ea1984}
C.~A. Phillips, J.~S. Petrofsky, D.~M. Hendershot, and D.~Stafford,
  ``Functional electrical exercise: A comprehensive approach for physical
  conditioning of the spinal cord injured patient,'' \emph{Orthopedics},
  vol.~7, no.~7, pp. 1112--1123, 1984.

\bibitem{Peng.Chen.ea2011}
C.-W. Peng, S.-C. Chen, C.-H. Lai, C.-J. Chen, C.-C. Chen, J.~Mizrahi, and
  Y.~Handa, ``Review: Clinical benefits of functional electrical stimulation
  cycling exercise for subjects with central neurological impairments,''
  \emph{J. Med. Biol. Eng.}, vol.~31, pp. 1--11, 2011.

\bibitem{Hunt.Fang.ea2012}
K.~J. Hunt, J.~Fang, J.~Saengsuwan, M.~Grob, and M.~Laubacher, ``On the
  efficiency of {FES} cycling: A framework and systematic review,''
  \emph{Technol. Health Care}, vol.~20, no.~5, pp. 395--422, 2012.

\bibitem{Hunt.Hosmann.ea2012}
K.~J. Hunt, D.~Hosmann, M.~Grob, and J.~Saengsuwan, ``Metabolic efficiency of
  volitional and electrically stimulated cycling in able-bodied subjects,''
  \emph{Med. Eng. Phys.}, vol.~35, no.~7, pp. 919--925, July 2013.

\bibitem{Szecsi.Krause.ea2007}
J.~Szecsi, P.~Krause, S.~Krafczyk, T.~Brandt, and A.~Straube, ``Functional
  output improvement in {FES} cycling by means of forced smooth pedaling,''
  \emph{Med. Sci. Sports Exerc.}, vol.~39, no.~5, pp. 764--780, May 2007.

\bibitem{Ibrahim.Gharooni.ea2008}
B.~S. K.~K. Ibrahim, S.~C. Gharooni, M.~O. Tokhi, and R.~Massoud,
  ``Energy-efficient {FES} cycling with quadriceps stimulation,'' in
  \emph{Proc. 13th Ann. Conf. of the Int. Funct. Electrical Stimulation Soc.},
  Freiburg, Germany, September 2008.

\bibitem{Gfohler.Lugner2000}
M.~Gf\"{o}hler and P.~Lugner, ``Cycling by means of functional electrical
  stimulation,'' \emph{IEEE Trans. Rehabil. Eng.}, vol.~8, no.~2, pp. 233--243,
  2000.

\bibitem{Idso.Johansen.ea2004}
E.~S. Ids{\o}, T.~Johansen, and K.~J. Hunt, ``Finding the metabolically optimal
  stimulation pattern for {FES}-cycling,'' in \emph{Proc. 9th Ann. Conf. of the
  Int. Funct. Electrical Stimulation Soc.}, Bournemouth, UK, September 2004.

\bibitem{Trumbower.Faghri2004}
R.~D. Trumbower and P.~D. Faghri, ``Improving pedal power during semireclined
  leg cycling,'' \emph{IEEE Eng. Med. Biol.}, vol.~23, no.~2, pp. 62--71,
  March-April 2004.

\bibitem{Hunt.Ferrario.ea2006}
K.~J. Hunt, C.~Ferrario, S.~Grant, B.~Stone, A.~N. McLean, M.~H. Fraser, and
  D.~B. Allan, ``Comparison of stimulation patterns for {FES}-cycling using
  measures of oxygen cost and stimulation cost,'' \emph{Med. Eng. Phys.},
  vol.~28, no.~7, pp. 710--718, September 2006.

\bibitem{Hakansson.Hull2009}
N.~A. Hakansson and M.~L. Hull, ``Muscle stimulation waveform timing patterns
  for upper and lower leg muscle groups to increase muscular endurance in
  functional electrical stimulation pedaling using a forward dynamic model,''
  \emph{IEEE Trans. Biomed. Eng.}, vol.~56, no.~9, pp. 2263--2270, September
  2009.

\bibitem{Perkins.Donaldson.ea2002}
T.~A. Perkins, N.~N. Donaldson, N.~A.~C. Hatcher, I.~D. Swain, and D.~E. Wood,
  ``Control of leg-powered paraplegic cycling using stimulation of the
  lumbro-sacral anterior spinal nerve roots,'' \emph{IEEE Trans. Neur. Sys. and
  Rehab. Eng.}, vol.~10, no.~3, pp. 158--164, September 2002.

\bibitem{Eser.Donaldson.ea2003}
P.~C. Eser, N.~Donaldson, H.~Knecht, and E.~Stussi, ``Influence of different
  stimulation frequencies on power output and fatigue during {FES}-cycling in
  recently injured {SCI} people,'' \emph{IEEE Trans. Neur. Sys. and Rehab.
  Eng.}, vol.~11, no.~3, pp. 236--240, September 2003.

\bibitem{Decker.Griffin.ea2010}
M.~Decker, L.~Griffin, L.~Abraham, and L.~Brandt, ``Alternating stimulation of
  synergistic muscles during functional electrical stimulation cycling improves
  endurance in persons with spinal cord injury,'' \emph{J. Electromyogr.
  Kinesiol.}, vol.~20, no.~6, pp. 1163 -- 1169, 2010.

\bibitem{Hakansson.Hull2012}
N.~A. Hakansson and M.~L. Hull, ``Can the efficacy of eletrically stimulated
  pedaling using a commercially available ergometer be improved by minimizing
  the muscle stress-time integral?'' \emph{Muscle Nerve}, vol.~45, no.~3, pp.
  393--402, March 2012.

\bibitem{Hunt.Stone.ea2004}
K.~J. Hunt, B.~Stone, N.-O. Neg{\aa}rd, T.~Schauer, M.~H. Fraser, A.~J.
  Cathcart, C.~Ferrario, S.~A. Ward, and S.~Grant, ``Control strategies for
  integration of electric motor assist and functional electrical stimulation in
  paraplegic cycling: Utility for exercise testing and mobile cycling,''
  \emph{IEEE Trans. Neur. Sys. and Rehab. Eng.}, vol.~12, no.~1, pp. 89--101,
  March 2004.

\bibitem{Chen.Yu.ea1997}
J.-J.~J. Chen, N.-Y. Yu, D.-G. Huang, B.-T. Ann, and G.-C. Chang, ``Applying
  fuzzy logic to control cycling movement induced by functional electrical
  stimulation,'' \emph{IEEE Trans. Neur. Sys. and Rehab. Eng.}, vol.~5, no.~2,
  pp. 158--169, June 1997.

\bibitem{Kim.Eom.ea2008}
C.-S. Kim, G.-M. Eom, K.~Hase, G.~Khang, G.-R. Tack, J.-H. Yi, and J.-H. Jun,
  ``Stimulation pattern-free control of {FES} cycling: Simulation study,''
  \emph{IEEE Trans. Syst. Man Cybern. Part C Appl. Rev.}, vol.~38, no.~1, pp.
  125--134, January 2008.

\bibitem{Farhoud.Erfanian2014}
A.~Farhoud and A.~Erfanian, ``Fully automatic control of paraplegic {FES}
  pedaling using higher-order sliding mode and fuzzy logic control,''
  \emph{IEEE Trans. Neur. Sys. and Rehab. Eng.}, vol.~PP, no.~99, p.~1, January
  2014.

\bibitem{Pons.Vaughan.ea1989}
D.~J. Pons, C.~L. Vaughan, and G.~G. Jaros, ``Cycling device powered by the
  electrically stimulated muscles of paraplegics,'' \emph{Med. Biol. Eng.
  Comput.}, vol.~27, pp. 1--7, 1989.

\bibitem{Petrofsky2003}
J.~S. Petrofsky, ``New algorithm to control a cycle ergometer using electrical
  stimulation,'' \emph{Med. Biol. Eng. Comput.}, vol.~41, no.~1, pp. 18--27,
  January 2003.

\bibitem{Liberzon2003}
D.~Liberzon, \emph{Switching in Systems and Control}.\hskip 1em plus 0.5em
  minus 0.4em\relax Birkhauser, 2003.

\bibitem{Hespanha.Liberzon.ea2008}
J.~P. Hespanha, D.~Liberzon, and A.~R. Teel, ``Lyapunov conditions for
  input-to-state stability of implusive systems,'' \emph{Automatica}, vol.~44,
  no.~11, pp. 2735--2744, November 2008.

\bibitem{Idso2002}
E.~S. Ids{\o}, ``Development of a mathematical model of a rider-tricycle
  system,'' Dept. of Engineering Cybernetics, NTNU, Tech. Rep., 2002.

\bibitem{Schutte.Rodgers.ea1993}
L.~M. Schutte, M.~M. Rodgers, F.~E. Zajac, and R.~M. Glaser, ``Improving the
  efficacy of electrical stimulation-induced leg cycle ergometry: An analysis
  based on a dynamic musculoskeletal model,'' \emph{IEEE Trans. Rehabil. Eng.},
  vol.~1, no.~2, pp. 109--125, 1993.

\bibitem{Sharma2009}
N.~Sharma, K.~Stegath, C.~M. Gregory, and W.~E. Dixon, ``Nonlinear
  neuromuscular electrical stimulation tracking control of a human limb,''
  \emph{IEEE Trans. Neural Syst. Rehabil. Eng.}, vol.~17, no.~6, pp. 576--584,
  2009.

\bibitem{Ferrarin2000}
M.~Ferrarin and A.~Pedotti, ``The relationship between electrical stimulus and
  joint torque: {A} dynamic model,'' \emph{IEEE Trans. Rehabil. Eng.}, vol.~8,
  no.~3, pp. 342--352, 2000.

\bibitem{Schauer2005}
T.~Schauer, N.~O. Negard, F.~Previdi, K.~J. Hunt, M.~H. Fraser, E.~Ferchland,
  and J.~Raisch, ``Online identification and nonlinear control of the
  electrically stimulated quadriceps muscle,'' \emph{Control Eng. Pract.},
  vol.~13, pp. 1207--1219, 2005.

\bibitem{Filippov1964}
A.~Filippov, ``Differential equations with discontinuous right-hand side,''
  \emph{Am. Math. Soc. Transl.}, vol. 42 no. 2, pp. 199--231, 1964.

\bibitem{Paden1987}
B.~Paden and S.~Sastry, ``A calculus for computing {F}ilippov's differential
  inclusion with application to the variable structure control of robot
  manipulators,'' \emph{IEEE Trans. Circuits Syst.}, vol. 34 no. 1, pp. 73--82,
  1987.

\bibitem{Chang1997}
G.-C. Chang, J.-J. Lub, G.-D. Liao, J.-S. Lai, C.-K. Cheng, B.-L. Kuo, and
  T.-S. Kuo, ``A neuro-control system for the knee joint position control with
  quadriceps stimulation,'' \emph{IEEE Trans. Rehabil. Eng.}, vol.~5, no.~1,
  pp. 2--11, Mar. 1997.

\bibitem{Kurosawa2005}
K.~Kurosawa, R.~Futami, T.~Watanabe, and N.~Hoshimiya, ``Joint angle control by
  {FES} using a feedback error learning controller,'' \emph{IEEE Trans. Neural
  Syst. Rehabil. Eng.}, vol.~13, pp. 359--371, 2005.

\bibitem{Jezernik2004}
S.~Jezernik, R.~Wassink, and T.~Keller, ``Sliding mode closed-loop control of
  {FES}: Controlling the shank movement,'' \emph{IEEE Trans. Biomed. Eng.},
  vol.~51, pp. 263--272, 2004.

\bibitem{Hausdorff1991}
J.~Hausdorff and W.~Durfee, ``Open-loop position control of the knee joint
  using electrical stimulation of the quadriceps and hamstrings,'' \emph{Med.
  Biol. Eng. Comput.}, vol.~29, pp. 269--280, 1991.

\end{thebibliography}

\end{document}